\documentclass[preprintnumbers,11pt,a4paper,oneside]{article}
\pdfoutput=1

\usepackage{contour}
\usepackage{xcolor}
\usepackage{float}
\usepackage{mathrsfs}

\contourlength{0.05em}

\usepackage{jheppub}
\usepackage{color}
\usepackage{graphicx}
\usepackage{wrapfig,enumerate,slashed}

\usepackage{footmisc}
\usepackage{amsmath}
\usepackage{wasysym} 
\usepackage{graphicx}
\usepackage{subcaption}
\usepackage{color}
\usepackage{comment}
\usepackage{appendix}
\usepackage{slashed}

\newcommand{\be}{\begin{equation}}
\newcommand{\ee}{\end{equation}}
\newcommand{\beq}{\begin{equation}}
\newcommand{\eeq}{\end{equation}}

\newcommand{\bee}{\begin{eqnarray}}
\newcommand{\eee}{\end{eqnarray}}
\newcommand{\beeq}{\begin{equation}}
\newcommand{\eeeq}{\end{equation}}

\newcommand{\ket}[1]{|#1 \rangle}

\usepackage{tikz}
\usepackage{orcidlink}

\makeatletter
\gdef\@fpheader{}
\makeatother
\begin{document}

\title{Real-Time Scattering Processes with Continuous-Variable Quantum Computers}

\author[a,b]{Steven~Abel\,\orcidlink{0000-0003-1213-907X}}
\author[a]{Michael Spannowsky\,\orcidlink{0000-0002-8362-0576}}
\author[a]{Simon Williams\,\orcidlink{0000-0001-8540-0780}}

\emailAdd{s.a.abel@durham.ac.uk}
\emailAdd{michael.spannowsky@durham.ac.uk}
\emailAdd{simon.j.williams@durham.ac.uk}

\affiliation[a]{\vspace{0.1cm} Institute for Particle Physics Phenomenology, Durham University, Durham DH1 3LE, UK}
\affiliation[b]{Department of Mathematical Sciences, Durham University, Durham DH1 3LE, UK}

\preprintA{IPPP/24/82}

\abstract{We propose a framework for simulating the real-time dynamics of quantum field theories (QFTs) using continuous-variable quantum computing (CVQC). Focusing on ($1+1$)-dimensional $\varphi^4$ scalar field theory, the approach employs the Hamiltonian formalism to map the theory onto a spatial lattice, with fields represented as quantum harmonic oscillators. Using measurement-based quantum computing, we implement non-Gaussian operations for CQVC platforms.
The study introduces methods for preparing initial states with specific momenta and simulating their evolution under the $\varphi^4$ Hamiltonian. Key quantum objects, such as two-point correlation functions, validate the framework against analytical solutions. Scattering simulations further illustrate how mass and coupling strength influence field dynamics and energy redistribution. Thus, we demonstrate CVQC’s scalability for larger lattice systems and its potential for simulating more complex field theories.}

\maketitle


\section{Introduction}

The Hamiltonian formalism offers a powerful framework for studying quantum field theories (QFTs), bypassing the so-called ``sign problem" which is prevalent in Monte Carlo methods for evaluating the path integral~\cite{PhysRevLett.46.77, VONDERLINDEN199253, PhysRevE.49.3855}. A key advantage of this approach is its ability to perform real-time evolution~\cite{doi:10.1126/science.1217069, 10.5555/2685155.2685163, Van_Damme_2021, PRXQuantum.3.020316, Papaefstathiou:2024zsu, Bennewitz:2024ixi, Rigobello:2021fxw, Abel:2024kuv, PhysRevResearch.6.033057, Jha:2024jan, Carena:2024peb, Ale:2024uxf, zemlevskiy2024, crane2024, Chai2025fermionicwavepacket}, enabling the study of non-equilibrium phenomena such as quantum quenches~\cite{Martinez:2016yna, PhysRevLett.109.175302, Ingoldby:2024fcy}, thermalisation~\cite{doi:10.1126/science.abl6277, PhysRevD.106.054508, PhysRevA.108.022612, Fromm:2023npm}, and dynamic phase transitions~\cite{Jensen:2022hyu, Angelides:2023noe, PRXQuantum.4.030323}. Importantly, this formalism is naturally compatible with quantum computing, where unitary transformations like $U(t) = e^{-iHt}$ enable efficient simulation of real-time dynamics in exponentially large Hilbert spaces using polynomial resources. Leveraging quantum computers for Hamiltonian simulation thus provides a pathway for studying the dynamics of complex quantum systems that are inaccessible to classical computation~\cite{Preskill:2018jim, Preskill:2018fag}.

In this work, we focus on the Hamiltonian formalism to explore the real-time dynamics of $\varphi^4$ scalar field theory in ($1+1$) dimensions using continuous-variable quantum computing (CVQC)~\cite{PhysRevLett.82.1784, Adesso2014, RevModPhys.77.513}. We reformulate the $\varphi^4$ theory in the Hamiltonian framework and discretise the spatial dimension into a lattice. While space is discretised, the field values at each lattice site are represented as quantum harmonic oscillators or qumodes. This mapping translates the field theory into a coupled oscillator system, making it well-suited for simulation on continuous-variable quantum computers. Additionally, CVQC benefits from practical hardware implementations, often based on photonic platforms~\cite{RevModPhys.79.135, 10.1063/1.5115814, Bromley_2020} that promise long coherence times and can be operated at room temperature, making it a promising approach for exploring quantum field theories and other problems involving continuous dynamics.

Our study focuses on developing methods to prepare initial quantum states corresponding to specific scattering-state configurations and implementing their real-time evolution under the $\varphi^4$ Hamiltonian. We capture the effects of field interactions by employing CVQC techniques, incorporating both Gaussian and non-Gaussian operations. Key observables, such as two-point correlation functions, are computed to investigate the dynamics of particle interactions and the propagation of field excitations. 
To validate our approach, we compare the results of CVQC simulations to analytical solutions of free scalar field theory, demonstrating the reliability and accuracy of our methods. Furthermore, we assess CVQC's scalability for simulating larger systems and longer-time evolutions. This approach can be extended straightforwardly to more complex and higher-dimensional quantum field theories.

In the following sections, we present the framework for the simulation of scalar field theories in the CVQC regime. In Sec.~\ref{sec:basics}, we outline the reformulation of the Hamiltonian and spatial discretisation onto a lattice of coupled Harmonic oscillators, and an efficient classical algorithm for approximating the CVQC simulation is presented. Section~\ref{sec:realTime} details the quantum algorithm and how the real-time evolution can be facilitated on a photonic device using experimentally realisable gate operations, emphasising techniques for handling both Gaussian and non-Gaussian operations. In Sec.~\ref{sec:phi4}, the model is validated by simulating $\varphi^4$ theory in (1+1) dimensions, including the preparation of scattering states, the computation of key observables such as correlation functions, and the simulation of a two particle scattering event. In Sec.~\ref{sec:aspects}, we discuss aspects of the quantum algorithm, specifically the preparation of scattering states on the device and the quantum resource scaling of the CVQC framework. Finally, in Sec.~\ref{sec:conclusions}, we offer a summary and conclusions. 

\section{Implementing field theory on a qumode lattice}\label{sec:basics}

In this section, we establish a dictionary between scalar quantum field theories and systems of coupled qumodes. A ``qumode'' can be considered as any quantum mechanical oscillator for this discussion. For the free-field theory, the qumodes in question will correspond to simple harmonic oscillators (SHOs). At the same time, non-linear systems will be quantum mechanical oscillators with more general potential wells. The oscillators we have in mind correspond to those found on continuous-time quantum computers. However, the dictionary is quite generic, and as we shall see, it will ultimately allow us to classically simulate QFT by coupling classically simulated qumodes as well.  

\subsection{Free fields on the qumode lattice}

Let us begin with the Lagrangian for the system we wish to encode, namely a scalar quantum field, $\varphi(x, t)$. To establish our dictionary, it is convenient to begin with a free field, where $\varphi$ has mass $\omega$ given in natural units ($\hbar=c=1$) such that 
\begin{equation}\label{eqn:Lagrangian} 
\mathscr{L} ~=~ \frac{1}{2} \left( \partial^\mu \varphi (x, t) \partial_\mu \varphi(x, t) - \omega^2 \varphi(x, t)^2\right)~,
\end{equation}
with $\varphi$ and $\pi$ obeying the usual equal-time commutation relations. 
The conjugate momentum associated with the field is  
\begin{equation}\label{eqn:conjMomenta}
\pi(x, t) ~=~ \frac{\partial \mathscr{L}}{\partial (\partial_0 \varphi(x, t))} ~=~ \partial_0 \varphi(x, t)~, 
\end{equation}
such that the Hamiltonian density is
\begin{equation}\label{eqn:HamiltonianDens} 
\mathscr{H}(x, t) ~=~ \frac{1}{2} \left( \pi(x, t)^2 + (\nabla \varphi(x, t))^2 + \omega^2 \varphi(x, t)^2 \right)~. 
\end{equation}

We wish to simulate this system by coupling together a lattice consisting of a large number, $N$, of quantum mechanical oscillators. For simplicity, we shall henceforth focus on a one-dimensional field theory and locate the oscillators at positions $x_n = an$ where $n\in [0,N-1]$, and where $a$ is the lattice spacing. We will also enforce periodic boundary conditions throughout, such that $x_{N}= x_0$, although this boundary condition will not be consequential in our discussion. 

Each quantum-mechanical oscillator has its own quadrature operators that we denote $\hat q_n$ and $\hat p_n$ (and can initially at least be assigned its own wavefunction $\psi_n(q_n)\equiv \langle q_n|\psi_n\rangle$). It is straightforward to verify that there is a configuration of nearest-neighbour couplings of the oscillators that reproduces the free-field theory Hamiltonian, $\mathscr{H}$, in the large $N$ limit. Moreover, the field operator $\varphi(x)$ and its corresponding conjugate momentum operator $\pi(x)$ can be directly equated with the quadrature variables of the oscillators (written as a  density) at that position on the lattice, such that one can identify  
\begin{equation}\label{eqn:latticeFields}
\hat q_n(t) ~\leftrightarrow ~  \varphi(x_n, t)~, \qquad \hat p_n (t) ~\leftrightarrow ~  \pi( x_n , t)~.
\end{equation}
To make this identification, we first express the spatial derivatives that appear in the Hamiltonian density of Eq.~\eqref{eqn:HamiltonianDens}  by finite difference,
\begin{equation}\label{eqn:finiteDifference}
\partial_i \varphi(x_n, t) ~\leftrightarrow~ \frac{{\hat q}_{n+1}(t) - {\hat q}_n(t)}{a}~.
\end{equation}
We then define the total Hamiltonian by summing the contributions from the coupled individual qumodes over the entire lattice as follows:
\begin{equation}\label{eqn:TotalHamiltonian}
H ~=~ \frac{a}{2} \sum_{n=0}^{N-1} \left[   {\hat p}_n(t)^2 + \left(\frac{\hat q_{\,\overline{n+1}}(t) - \hat q_n(t)}{a}\right)^2 +  \omega^2 \hat q_n(t)^2 \right]~,
\end{equation}
with variables normalized such that 
\beq 
[ \hat q_n ,\,\hat p_m]~=~ i\delta_{nm} ~,
\eeq
and where periodic boundary conditions have been enforced using 
\begin{equation}
    \overline{n + 1} ~=~ n+1 \mod (N)~. 
\end{equation}
The $a$ prefactor in the Hamiltonian ensures that it scales extensively with the size of the lattice, ${\rm Vol}=L=aN$. We can also define the combined wavefunction and the combined set of quadrature variables, 
\beq 
\label{eq:combined}
 |\Psi \rangle ~=~ \otimes _n |\psi_n\rangle ~~;~~~ |{\mathbf q}\rangle  ~=~ \otimes _n |q_n\rangle~.
 \eeq
 In the ground state of the lattice, all the wave functions will be degenerate, and there will be no contribution to the Hamiltonian from the gradient terms:
\beq \langle {\mathbf q} | \left(\frac{\hat q_{\,\overline{n+1}}(t) - \hat q_n(t)}{a}\right)^2 | \Psi_{\rm degen} \rangle ~=~ 0~.\eeq
Evidently, $\omega$ is the natural frequency of the lattice of oscillators when the qumodes are all degenerate. The ground state energy of the system is then simply the sum of the ground-state energies of all the qumodes, $H_0 = \frac{\omega}{2}\,aN$, which is equivalently the volume multiplying the ground-state energy of a single qumode.
Perturbing this vacuum in a degenerate manner would result in a qumode lattice wavefunction  that continued to factorise into a product of independent qumodes, 
\beq \label{eq:unentangled}\langle {\mathbf q} |  \Psi_{\rm degen} \rangle ~=~ \prod_n \langle q_n |\psi_n \rangle ~.\eeq
Of course, propagating modes then correspond to perturbations of the lattice around such a degenerate ground state. However it is important to realise that a lattice of qumodes in their ground state does {\it not} correspond to the traditional QFT vacuum,  precisely due to the spatial kinetic terms, but it effectively coincides with the QFT vacuum for non-relativistic modes. We will return to the question of the QFT vacuum in detail later.

To make explicit the relation anticipated in Eq.~\eqref{eqn:latticeFields}, we must first diagonalise $H$. This is made non-trivial by the hopping terms which provide a ``circulant'' contribution. Defining 
$\nu = e^{2\pi i /N}$, the diagonalisation matrix is found to be 
\beq 
U_{\alpha m}~=~ \frac{1}{\sqrt{N} } \nu^{\alpha m}~,
\eeq
where we use Greek letters $\alpha, \beta, \ldots , \kappa $ to refer to the diagonal basis. In this diagonal basis, the Hamiltonian of Eq.~\eqref{eqn:TotalHamiltonian} becomes, 
\begin{equation}
H ~=~ \frac{a}{2} \sum^{N-1}_{\alpha =0} \left( \vert \hat{p}_\alpha \vert ^2 + \omega_\alpha ^2 \vert \hat q_\alpha  \vert ^2 \right)~,\\[1em]
\end{equation}
where \beq 
\label{eq:rotations}
\hat {p}_\alpha ~=~ \sum_n U_{\alpha n} \hat p_n~~;~~~~
\hat {q}_\alpha ~=~ \sum_n U_{\alpha n} \hat q_n~,
\eeq
and where the eigenvalues are
\begin{equation}\label{eqn:omegas}
\omega_\alpha ~=~ \sqrt{\left( \omega^2 + \frac{4}{a^2}\sin^2 \left( \frac{\pi \alpha }{N} \right) \right)}~.
\end{equation}
Thus, in the $N\to \infty $ limit we recover the usual interpretation of free-field QFT as a tower of SHO oscillators, which have momentum 
\beq 
\label{eq:NinftyK}
k_\alpha ~=~ 2\pi \frac{\alpha}{aN}~=~ 2\pi \frac{\alpha}{L}~,
\eeq
and which obey the relativistic relation $\omega_\alpha^2=\omega^2 +k_\alpha^2$.

Two remarks are in order. First note that in the diagonal basis, $\hat q_\alpha$ and $\hat p_\alpha $ are complex, with a phase given by the momentum $k_\alpha$, despite the fact that $\hat q_n$ and $\hat p_n$ are real. Consequently, each real mode in space is now represented by a {\it pair} of complex conjugate modes in momentum (Fourier) space, corresponding to left and right-moving momentum eigenstates. As their sum must be real, we can conclude that $\hat q_\alpha $ can only be considered independently observable in conjunction with momentum conservation in the QFT.

Our second remark is that it is clear from Eq.~\eqref{eqn:omegas} that the special relativistic relation $\omega _\alpha =\sqrt{ \omega^2 + k_\alpha^2 }$ is recovered only under the assumption that $\pi \alpha/N= a k_\alpha/2 \ll 1$. We can thus anticipate the possibility of non-relativistic artefacts on a finite lattice when the momenta approach unity in units of $a^{-1}$. As the lattice momentum is really $\tilde k_\alpha = (2/a) \sin ( \pi \alpha/N) $ these artifacts would resemble superluminal states that satisfy $\sin^{-1} (a \tilde k_\alpha/2) \approx a k_\alpha/2$, i.e. they would have $\alpha' = \alpha+ 2 \beta N $ where $\beta  \in {\mathbb Z}$.
Likewise, the lattice's highest physical momentum can be described as $k_\alpha \sim 1/a $. Hence the lattice-spacing $a$ defines a natural UV cut-off, and relativistic behaviour can only be modelled for modes that have $\omega a \ll 1$.

We are now able to make the direct identification promised in Eq.~\eqref{eqn:latticeFields} between $\varphi(x_n)$ and the quadrature variable at that lattice site $q_n$. To do this, recall that the ladder operators of the qumodes on the lattice are 
\begin{align}\label{eqn:ladderOps}
\hat a_n &~=~ \sqrt{\frac{\omega  }{2}} \left( \hat q_n + \frac{i}{\omega } \hat p_n\right)~, \nonumber\\
\hat a^\dagger_n &~=~ \sqrt{\frac{\omega   }{2}} \left( \hat q_n - \frac{i}{\omega } \hat p_n\right)~,
\end{align} 
normalized such that $H = \sum_n a \omega (\hat a^\dagger _n \hat a_n +\frac{1}{2})$ when all the qumodes are degenerate.
In the diagonal basis these become \begin{align}\label{eqn:momLadders}
\hat{a}_\alpha 			&~=~ \sqrt{\frac{\omega_\alpha  }{2 }}~\sum_n\left(  U_{\alpha n}\,\hat q_n + \frac{i}{\omega_\alpha} U_{\alpha n}\,\hat p_n \right)~,\nonumber\\
\hat{a}_\alpha^\dagger 	&~=~ \sqrt{\frac{\omega_\alpha }{2 }}~\sum_n\left( \hat q_n \left(U^\dagger\right)_{n\alpha} - \frac{i}{\omega_\alpha} \hat p_n \left(U^\dagger\right)_{n\alpha} \right)~.
\end{align}
The fact that the field operator at position $x_n$ is simply the quadrature variable at that point on the lattice can be seen as follows: our claim is that
\begin{align}
\varphi(x_n) &~=~ \hat q_n ~=~ \frac{1}{\sqrt{2\omega }  } (\hat a_n + \hat a_n^\dagger ) \nonumber \\
 &~=~ \sum_\alpha  \frac{1}{\sqrt{2\omega_\alpha } }\left( U_{n\alpha}^\dagger \hat{a}_\alpha  + \hat{a}^\dagger_\alpha U_{\alpha n}  \right)~,\nonumber \\
 &~=~  \sum_\alpha \frac{1}{\sqrt{2\omega_\alpha \,N}}\left( \hat{a}_\alpha e^{-i\alpha n} + \hat{a}^\dagger_l e^{i \alpha n} \right)~,\nonumber \\ 
 &~=~ \sum_\alpha \frac{1}{\sqrt{2\omega_\alpha\,N }}\left( \hat{a}_\alpha e^{-ik_\alpha x_n} + \hat{a}^\dagger_\alpha e^{ik_\alpha x_n} \right)~,
\label{eq:phi_to_q} \end{align}
where we use the $N\to \infty $ definition of the momentum in Eq.~\eqref{eq:NinftyK}. Similarly, its conjugate momentum is
 \begin{align}
\pi (x_n)   &~=~ \hat p _n ~=~  \sum_\alpha (-i) \sqrt{\frac{\omega_\alpha}{2 a N }} \left ( \hat{a}_\alpha e^{-ik_\alpha x_n} - \hat{a}^\dagger_\alpha e^{ik_\alpha x_n} \right)~.
\label{eq:pi_to_p}
\end{align}
Taking the continuum limit as $N\rightarrow \infty$, and replacing $\sum_\alpha \equiv \frac{aN}{2\pi } \int dk$ we retrieve the standard QFT definition of the field as an expansion in momentum space at a fixed time, namely 
\begin{align}
\varphi(x) &~=~ \int \frac{\textrm{d} k}{(2\pi)} \frac{1}{\sqrt{2\omega_k}}\left( \hat{a}_k e^{ikx} + \hat{a}^\dagger_k e^{-ikx} \right)~,\nonumber \\[1em]
\pi (x)   &~=~  \int \frac{\textrm{d} k}{(2\pi)} (-i) \sqrt{\frac{\omega_k}{2}} \left ( \hat a_k e^{ikx} - \hat a^\dagger_k e^{-ikx} \right)~,
\end{align}
where $\omega_k=\sqrt{\omega^2 + k^2}$, and where we have made the usual rescaling of the ladder operators by the square-root of the volume, namely $\sqrt{a} \,\hat a_\alpha \longrightarrow  \hat a_k /{\sqrt{a N}}$, in order to be consistent with  $[\hat a_k,\hat a_{k'}^\dagger ]=2\pi a \delta(k-k')$, versus   
$[\hat a_\alpha,\hat a_{\beta}^\dagger ]= \delta_{\alpha\beta}$.

\subsection{Hamiltonian evolution for interacting theories and Trotterization}

So far we have established an exact correspondence between scalar free field theory and a coupled lattice of SHO qumodes in the infinite lattice limit. 
 However, once an interacting (i.e. non-quadratic) potential is introduced, 
\begin{equation}
\label{eq:vvv}
    \mathscr{V}(\varphi)~=~ \mathscr{V}_I(\varphi)~+~ \frac{1}{2} 
    {\omega^2 \varphi^2}~,
\end{equation}
the two procedures part company. In the second-quantized QFT, one abandons any hope of including the influence of this extra contribution to the potential on the {\it definition} of the scattering states themselves. In other words, the asymptotic in-states and out-states are always considered to be approximately ``free-field" excitations, defined by the action of the SHO ladder operators $\hat a_\alpha$ and $\hat a^\dagger_\alpha$ on the free-field vacuum. Perturbation theory is then used to expand around this free-field approximation and determine how $\mathscr{V}_I$ couples the fields to each other in a central interaction region. 

In contrast, in the Hamiltonian lattice approach, even when $\mathscr{V}_I$ is introduced, one retains the full first-quantized lattice, which ensures that propagating modes are always built out of excitations of the actual vacuum of the theory as defined by the {\it full}  potential $V$. In fact, every oscillator on the lattice retains all its local quantum mechanical information about how the potential $\mathscr{V}$ modifies the spectrum of states.  In principle, it is even possible for a Hamiltonian lattice to incorporate potentials that destabilize the vacuum locally and allow non-perturbative tunnelling processes.

The great advantage of the Hamiltonian lattice approach is that in principle one can naturally simulate the real-time evolution of a QFT via the Schr\"odinger time-evolution operator,
\begin{equation}
     \mathcal{U}(t) ~=~ e^{-iHt}~,
\end{equation}
which propagates an initial state of the entire lattice $\ket{\Psi(t=0)}$ to the state $\ket{\Psi(t)} = \mathcal{U}(t) \ket{\Psi(0)}$ at time $t$. Simulating real-time evolution in this manner avoids the sign problem which is ubiquitous in the path-integral formalism~\cite{PhysRevLett.46.77, VONDERLINDEN199253, PhysRevE.49.3855}.

Unfortunately, directly computing the evolution is typically infeasible for large times because the evolution operator $\mathcal U$  becomes exponentially large and is, in general, dense. Because of this, one typically instead approximates the time-evolution operator by employing Trotter-Suzuki decomposition~\cite{pjm/1103039709, 10.1063/1.526596}, such that 
\begin{equation}
    \mathcal{U}(t) ~=~ \left[\prod_i e^{-iH_i \delta t} \right]^{t/\delta t} + \mathcal{O}\left( \delta t^2\right)~, 
\end{equation}
where the Hamiltonian has been decomposed into a sum of non-commuting parts, $H=\sum_i H_i$. This approximates the time-evolution operator $\mathcal{U}(t)$ up to an error of $\mathcal{O}\left(\delta t^2\right)$, and thus is a good approximation for $\delta t \ll 1$. We will refer to this process as \textit{Trotterisation} and it is the method that we will utilise here. 

One of the benefits of Trotterisation in the context of QFT is that the Hamiltonian itself is not dense due to the locality of the Hamiltonian density. Thus, as we shall see, Trotterised evolution can in principle be easily implemented in a
quantum circuit of qumodes that have the correct corresponding couplings. 
We will describe the relevant circuits in Sec.~\ref{sec:realTime}. 

\subsubsection{General qumode lattice structure, and classical simulation}

However, to build towards the quantum qumode circuit, we now note that the Trotterised evolution is sufficiently simple that it can also be simulated using a lattice of coupled ``classical modes'' with the Trotterised evolution being executed numerically. We briefly develop that formalism for the remainder of this section because it will inform the quantum circuit required for the quantum lattice. In particular, it will make manifest the advantage of the full quantum lattice.

To implement a ``classical lattice of modes'', consider the Hamiltonian density of scalar QFT in the continuum in its most general form with an arbitrary interaction, $\mathscr{V}_I$, such that
\begin{equation}
    \mathscr{H}(x, t) = \frac{1}{2} \left( \pi(x, t)^2 + \left(\nabla \varphi(x, t)\right)^2 + \omega^2\varphi(x, t)^2 \right) + \mathscr{V}_I(\varphi(x, t))~.
\end{equation}
Following the same discretisation procedure outlined above for the free field theory, on a lattice  of $N$ sites the total Hamiltonian for this system becomes,
\begin{equation}
    H ~=~ a\sum_n \left[ \frac{1}{2} \left(\hat{p}_n(t)^2+ \left(\frac{\hat{q}_{\,\overline{n+1}}(t) - \hat{q}_n(t)}{a}\right)^2 + \omega^2\hat{q}_n(t)^2 \right) + \mathscr{V}_I\left(\hat{q}_n\right(t)) \right]~.
\end{equation}
Expanding the terms within the Hamiltonian, 
we find three kinds of contribution in the total Hamiltonian:
\begin{equation}\label{eqn:CVQCHam}
    H a^{-1} =  \sum_n \left[ \frac{1}{2} \left(\hat{p}_n(t)^2 + \omega^2\hat{q}_n(t)^2\right) + V_I(\hat{q}_n) \right] - \frac{1}{a^2} \sum_n \hat{q}_{\,\overline{n+1}}\hat{q}_n~,
\end{equation}
where $V_I$ acts as an effective potential \begin{equation}\label{eqn:effectivePot}
    V_I ~=~ \frac{1}{a^2} \hat{q}_n^2 + \mathscr{V}_I(\hat{q}_n)~.
\end{equation}
We will henceforth absorb the overall factor of $a^{-1}$ into our definition of $\delta t$, which can, therefore, be thought of as rescaling the time in lattice units. The first term in $H$ is simply the sum of the Gaussian SHO piece of every qumode. This piece is diagonal but does not commute with the others because it contains $\hat p_n$. The term $V_I$ is the remaining diagonal contribution coming from the non-Gaussian interactions, as well as the diagonal quadratic term from the gradients in the Hamiltonian density. Meanwhile, the ``hopping-terms'' provide the same nearest-neighbour interactions that we met previously in the free-field theory and are the only pieces that make the lattice non-trivial. 

Thus, Eq.~\eqref{eqn:CVQCHam} highlights a remarkable simplicity inherent in the qumode formulation of QFT. Regardless of the potential $\mathscr{V}_I$, the Hamiltonian always ultimately reduces to a direct sum of these three terms, namely the SHO,  an arbitrary potential $V_I$, and finally a ring of ``hopping-term" interactions. Consequently, to simulate the real-time evolution of a scalar QFT, the crux of the problem is to implement these three stages of quantum-mechanical time-evolution within each Trotter step. 

For practical purposes, it is extremely useful to consider the first two diagonal stages of the Trotter step in isolation. This is because, as we have already seen, the kinetic terms cancel exactly when the qumodes are degenerate on the lattice. Thus, coherent oscillation of the interacting theory is trivial, in the sense that it is isomorphic to the oscillation of a single qumode with interaction $V_I=\mathscr{V}_I$. This provides a useful physical situation for testing the interacting theory where the hopping terms are guaranteed to be inert. 

Focussing on these diagonal parts of the Trotter evolution for the moment, they act on a single qumode (with quadrature variables $\hat q$ and $\hat p$) as follows:
\beq
\mathcal{U}_{\rm diag}(\delta t) ~=~ \mathcal{U}_R(-\delta t) ~\mathcal{U}_V(\delta t) ~,
\eeq
where
\beq 
 \mathcal{U}_R(-\delta t)~=~ \exp\left(-i \frac{1}{2} (\hat p^2+\omega^2 \hat q^2) \delta t\right) ~
\eeq
contains the SHO contribution, which in the language of continuous-variable quantum computing would be called a ``rotation gate'', and where 
\beq 
 \mathcal{U}_V(-\delta t)~=~ \exp\left(-i V_I \mathscr(\hat q) \delta t\right) ~
\eeq
would be a so-called ``non-Gaussian gate''.

To implement the action of these gates it is convenient for the classical simulation to (at the risk of confusion) discretise the quadrature variable $q$. 
That is we define 
\beq\label{eq:discretise}
q_j ~=~ j\, \xi  - L/2 ~,
\eeq 
where  $j\in [0,M-1]$ and $\xi = L/M$ where the interval is $L$.
Now the action of the $\mathcal{U}_R$  gate on a single qumode can be implemented by making a return visit to the Fock basis:
\begin{align}
\langle q_i |\mathcal{U}_R(-\delta t) | \psi\rangle ~&=~  
\sum_\ell \langle q_i | \ell \rangle  e^{ -i\delta t \omega (\ell+\frac{1}{2} )} \langle \ell  | \psi\rangle \nonumber \\
~&=~  
\xi \sum_j  \sum_\ell \langle q_i | \ell \rangle  e^{ -i\delta t \omega (\ell+\frac{1}{2} )} \langle \ell  | q_j\rangle \langle q_j | \psi\rangle~.
\end{align}
Thus the rotation gate acts on the wavefunction as 
\beq 
\langle q_i |\mathcal{U}_R(-\delta t) | \psi\rangle ~=~ \sum_j \mathcal{U}_{R,ij}\,  \langle q_j  | \psi\rangle~,
\eeq 
where we can approximate all the moments in advance by using a truncated sum over the Fock basis: 
\beq 
\mathcal{U}_{R,ij} ~=~ \xi \sum_{\ell=0}^{\ell_{\rm trunc}} \,\langle q_i | \ell \rangle \, e^{ -i\delta t \omega (\ell+\frac{1}{2} )} \, \langle \ell  | q_j\rangle ~.
\eeq 
It is typically sufficient to take $\ell_{\rm trunc} = 80$. Meanwhile, the non-Gaussian gate acts on the wavefunction as 
\beq 
\langle q_i |\mathcal{U}_V(-\delta t) | \psi\rangle ~=~ \sum_j \mathcal{U}_{V,ij} \, \langle q_j  | \psi\rangle~,
\eeq 
where the matrix $\mathcal{U}_{V,ij}$ is (for consistent confusion) given by
\beq 
\mathcal{U}_{V,ij} ~=~ \delta_{ij} \, 
e^{-i \delta t V_I(q_i) } ~.
\eeq 

\begin{figure}[t!]
\centering
\includegraphics[keepaspectratio, width=0.75\textwidth]{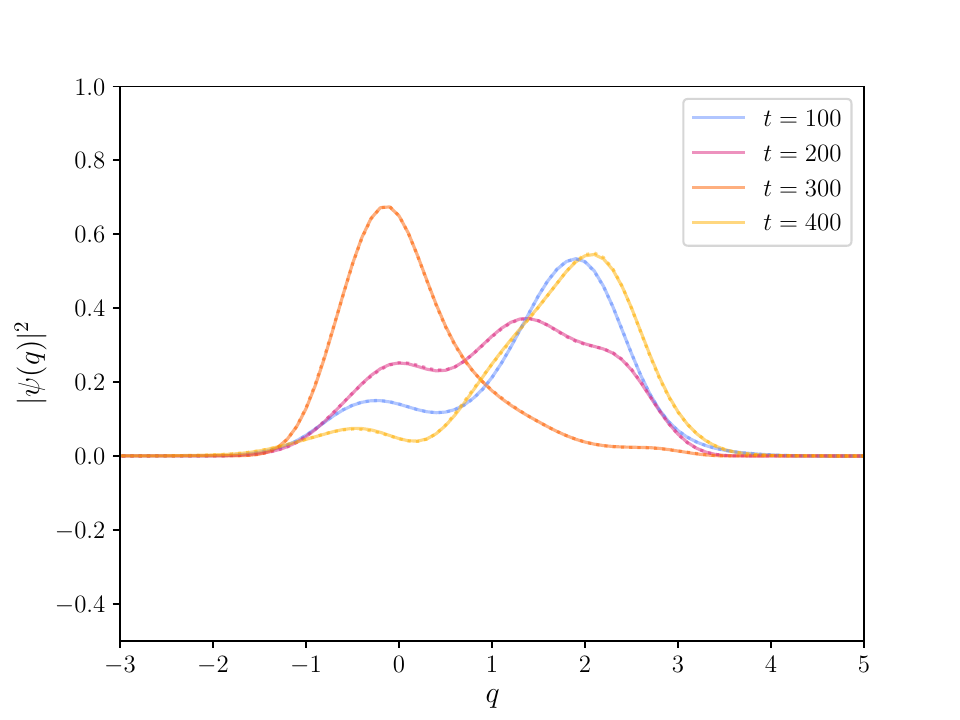}
\caption{Trotterised evolution of a single qumode under the potential in Eq.~\eqref{eq:pot_test} (solid lines) versus the wave-function evolved using Qibo~\cite{qibo_paper} (dotted). The Trotter time-step is $\delta t=0.01$. Thus, the evolution represents $(1 - 4) \times 10^4$ Trotter steps, respectively, for the four plots.}
\label{fig:testplot}
\end{figure}

Now let us as promised test the accuracy of the Trotterisation procedure by testing these diagonal contributions to the evolution against a toy potential: in this case, we can consider the same potential as in Ref.~\cite{Abel:2024kuv}, namely 
\beq 
\label{eq:pot_test}
V_I ~=~ - \frac{(1+\epsilon/4)}{2}  q^3 + \frac{1}{8} q^4~, 
\eeq 
where $\epsilon = 0.1$, with $\omega=1$.  (Note that the diagonal $\hat q^2/a^2$ piece is absent here by choice, simply to make a direct comparison.)
An example of the evolution with $\delta t=0.01$ is shown in Fig.~\ref{fig:testplot} (as solid lines) and compared to the exact evolution obtained using Qibo \cite{qibo_paper} (shown using dotted lines). In these plots, we discretise the quadrature $q$ as described above, with $M=200$. The agreement is remarkable over thousands of Trotter steps. On the complete lattice, this procedure and the subsequent evolution are valid if we assume that the qumodes are initially prepared un-entangled so that they can be written as a simple product of single qumode wavefunctions as in Eq.~\eqref{eq:unentangled}.

Finally, we come to the non-diagonal contribution to the QFT evolution which arises from the hopping term; namely we are required to also consider the Hubbard-like evolution operator 
\begin{equation}\label{eq:qhop}
\mathcal{U}_\textrm{hop} ~=~ \prod_{n=1}^N  e^{i a^{-2} \, \hat{q}\,_{\overline{n+1}} \,\hat{q}_n \, \delta t}~
\end{equation}
acting on the lattice. 

In the classical simulation, incorporating this operator is non-trivial and requires some care about the kinds of approximations we are making, because the operator will entangle the qumodes, and this entanglement will itself propagate through the lattice. We should stress at this point that this difficulty is entirely a consequence of the classical evolution that we are using to simulate the lattice, and as we shall see, it is at this point that the real quantum lattice will show its great advantage. 

To see how to proceed in the classical simulation, let us consider the starting state of the lattice a bit more carefully, in particular the nature of the vacuum in QFT. As we already mentioned, it is convenient to assume that the wavefunctions of the qumodes are not initially entangled, so that they can be written as a product of wavefunctions over the qumodes. If the initial state is a perturbation of the vacuum in which every qumode is in its SHO groundstate, then in the quadrature basis the vacuum is 
\beq \label{eq:qumodeGroundState}
\langle {\bf q}  |{\bf 0} \rangle ~=~ \prod_{n = 1}^N \langle q_n |0_\omega \rangle~,
\eeq
which is a product of Gaussian SHO wavefunctions and hence proportional to 
\beq \langle {\bf q}  |{\bf 0} \rangle ~\propto ~ \exp \left(
- \omega \sum_n q_n^2\right)~.
\eeq

However, this is as we already mentioned {\it not} the vacuum of the QFT. The vacuum of the QFT is obtained by solving $\hat a_\alpha |{\bf 0}\rangle =0$ for all $\alpha$. It, therefore, takes a Gaussian form that is not factorised, but instead given by 
\beq \label{eq:trueVac}
\langle {\bf q}  |{\bf 0} \rangle_{\rm QFT} ~=~ \prod_{\alpha = 0}^{N-1} \langle q_\alpha |0_{\omega_\alpha} \rangle~,
\eeq
which is proportional to\footnote{For completeness, in the continuum limit this becomes 
\beq 
\langle {\bf q}  |{\bf 0}  \rangle_{\rm QFT} ~ \propto ~ \exp \left(
- \int \frac{dk} {2\pi} \, \frac{\omega_k}{2} \, |\tilde{\phi}(k)|^2\right) ~,
\eeq
where ${\tilde \phi} (k)$ is the Fourier transform of $\phi$.}
\begin{align} 
\langle {\bf q}  |{\bf 0}  \rangle_{\rm QFT} ~&\propto ~ \exp \left(
- \sum_\alpha \omega_\alpha  |q_\alpha |^2/2\right)~\nonumber \\
~&= ~ \exp \left(
- \sum_\alpha \omega_\alpha  \sum_{nm} U_{\alpha n} U^\dagger_{m\alpha} q_n q_m /2\right) ~.
\end{align}
Indeed it is straightforward to verify from Eq.~\eqref{eqn:momLadders} that 
\begin{align}
\langle {\bf q}   | \hat a_\alpha |{\bf 0}\rangle _{\rm QFT} ~& \propto~ \sum_n U_{\alpha n} q_n - \frac{1}{\omega_\alpha }\sum_\beta 
\frac{\omega_\beta }{2}\sum_{n m }  U_{\alpha n} ( U_{\beta n} U^\dagger_{m\beta} + U_{\beta m} U^\dagger_{n\beta}) \,q_m  \nonumber \\
&= ~ \sum_n U_{\alpha n} q_n -  
\frac{1 }{2}\sum_{m}  (  U^\dagger_{m(-\alpha) } + U_{\alpha m} )\, q_m   
~=~0~,
\end{align}
where we use, $U^\dagger_{m(-\alpha) } = U_{\alpha m }$.
In contrast with the completely factorised vacuum that we wrote above, the QFT vacuum appears to maximally entangle the qumodes of the lattice, although the vacuum is, of course, diagonal in the momentum basis. Thus, this apparent entanglement is only measurable when a) we excite high momentum modes and b) the theory is interacting (the latter condition arising because it is specifically the non-Gaussian interactions which favour the qumode basis). In other words if we restrict the discussion to non-relativistic modes where $\omega_\alpha \approx \omega $ then we may use the approximation  
\beq 
\sum_\alpha \omega_\alpha  \sum_{nm} U_{\alpha n} U^\dagger_{m\alpha} q_n q_m /2  ~\approx ~ \sum_\alpha \omega \sum_{nm} U_{\alpha n} U^\dagger _{m \alpha }q_n q_m /2 ~=~ \omega \sum_n q_n^2
\eeq
to write 
\beq \label{eq:nonRelApprox}
\langle {\bf q}  |{\bf 0} \rangle_{\rm QFT}~\simeq ~ \langle {\bf q}  |{\bf 0} \rangle ~.
\eeq
As long as we only excite non-relativistic modes, the difference in the definition of the vacua simply adds an unmeasurable time-independent phase to the wavefunction which can be ignored. 
Therefore, the restriction to non-relativistic modes is convenient for state preparation on a qumode lattice. We will discuss later how a genuine QFT vacuum could be prepared, but for our simulations we will make this simplification and consider perturbations of the decoupled lattice vacuum rather than the QFT vacuum. (It is probably worth recalling that there is also a natural momentum limit $k_\alpha \lesssim 1/a$, as the sum over $\alpha$ that defines the QFT vacuum on the lattice is truncated.) 

This approximation implies a general condition on the initial states that we are allowed to consider: they must be built from low momentum excitations of the vacuum,
\beq
\label{eq:initPsi}
|{\bf \Psi} (0) \rangle ~=~ \sum_\alpha c_\alpha \hat q_\alpha |{\bf 0} \rangle~,
\eeq
where $c_\alpha $ are some coefficients that obey the above non-relativistic criterion, namely that $c_\alpha \ll 1 $ for $\alpha/N \geq a\omega /2\pi$.
This is the usual assumption for non-relativistic scattering states in QFT, and it is certainly the case if, for example, we prepare the lattice so that there is an incoming non-relativistic Gaussian wavepacket state, which we will define explicitly later. 

Under the above assumptions, we may approximate the effect of Hubbard-like hopping terms by entanglement truncation as follows. We can imagine after each Trotter step, which involves the hopping terms, redefining our qumodes by projecting the new state $\ket{\Psi'}= \exp (-i \sum_n  q_n q_{\overline{n+1}}  \delta t/a^2 ) \ket{\Psi}$ onto the quadrature variables. It is straightforward to show that this is equivalent to the following transformation on the basis of qumode wavefunctions: 
\begin{align}
\langle q | \psi'_n \rangle ~=~  
\sum_m~ \widetilde{\mathcal U}_{{\rm hop} , nm} \, \langle q | \psi_m \rangle   
\end{align} 
where 
\begin{align}
\widetilde{\mathcal U}_{{\rm hop} , nm} &~=~ \sum_{\alpha} U_{n\alpha }^\dagger  \, e^{i \delta t \cos(2\pi \alpha/N) \,{q}^2/a^2} 
\,  U_{\alpha  m}~.
\end{align} 
In performing such a step we are truncating the degree of entanglement to neighbouring qumodes as in TEBD~\cite{Vidal_2003,Vidal:2003lvx}.

The resulting combined Trotter-Suzuki transformation of the qumodes, 
\beq 
\ket {\psi_n} ~\to~ \sum _m ({\mathcal U}_R\,\, {\mathcal U}_{V} \,\,  \widetilde {\mathcal U}_{\rm hop})_{nm} \,\, \ket {\psi_m}~,
\eeq
will be the basis for our classical simulation of the qumode lattice. This approximation will be useful to test our ideas about the quantum lattice; however, as we have stressed and as we are about to see, a genuine quantum lattice would not require any approximation of ${\mathcal U}_{\rm hop}$, and the qumodes could become maximally entangled without presenting any particular difficulty. 

\section{Real-time evolution on continuous-variable quantum computers}\label{sec:realTime}

Therefore, let us now turn to the implementation of a qumode lattice on real physical devices. For this purpose, we will consider quantum computation via quantum optics, which offers an experimentally realisable method of continuous-variable quantum computing (CVQC) using the infinite-dimensional photon-number degree of freedom~\cite{RevModPhys.77.513, Adesso2014}. Photonic devices benefit from the exceedingly low decoherence characteristics of photons, enabling quantum information to be maintained and transmitted through numerous quantum gate operations~\cite{RevModPhys.79.135, Bromley_2020}. Consequently, photonic devices are particularly well suited to the real-time simulation of quantum field theories, which are often limited due to decoherence effects in quantum devices over long evolution times. 

As explained in Sec.~\ref{sec:basics}, a set of $N$-qumodes can be modelled as a set of $N$-quantum oscillators. On a typical CVQC device the qumodes are SHOs with the Hamiltonian,  
\begin{equation}\label{eqn:SHO}
H_\textrm{SHO} = \frac{1}{2} \left( \hat{p}^2 + \omega^2 \hat{q}^2 \right)~,
\end{equation}
which can easily reproduce the Gaussian rotation gates ${\mathcal U}_R$. This would be a good starting point for the free-field QFT, however for interacting systems the qumodes in the lattice are anharmonic with a non-Gaussian potential, with the Hamiltonian taking the form of the first term in Eq.~\eqref{eqn:CVQCHam}.

A procedure for efficiently performing the time evolution of a quantum mechanical oscillator state with an arbitrary potential using an experimentally realisable quantum gate set was formulated in~\cite{Abel:2024kuv}. Thus, it is already possible to realise the diagonal gates, ${\mathcal {U}}_R$ and ${\mathcal {U}}_V$, on CVQC devices. This is very convenient because it allows one to build up the QFT in a modular fashion. That is, first one can focus on simulating the real-time quantum-mechanical evolution of a single qumode wavefunction, corresponding to a single lattice site.  With this to hand one can then build up the QFT by connecting qumodes together with the ${\mathcal U}_{\rm hop}$ hopping terms of Eq.~\eqref{eq:qhop}.

Therefore, let us first consider simulating the real-time evolution of a single anharmonic oscillator qumode on a CVQC device. As described in~\cite{Abel:2024kuv}, one of the main challenges in photonic quantum computing is the lack of gate operations that induce strong non-Gaussian effects, which are necessary to achieve universal quantum computation~\cite{PhysRevLett.82.1784, PhysRevA.100.012326, PhysRevA.100.052301}, and which in this particular context are implied by the ${\mathcal U}_V$ operator. To do this we will closely follow the approach outlined in Ref.~\cite{Abel:2024kuv}, which realised the evolution of a single qumode in an arbitrary potential through a measurement-based machine-learning-enhanced scheme~\cite{PhysRevA.100.012326, PhysRevA.100.052301}.

\begin{figure}[t]
\centering
\begin{subfigure}[c]{0.9\textwidth}
\centering
\includegraphics[keepaspectratio, width=\textwidth]{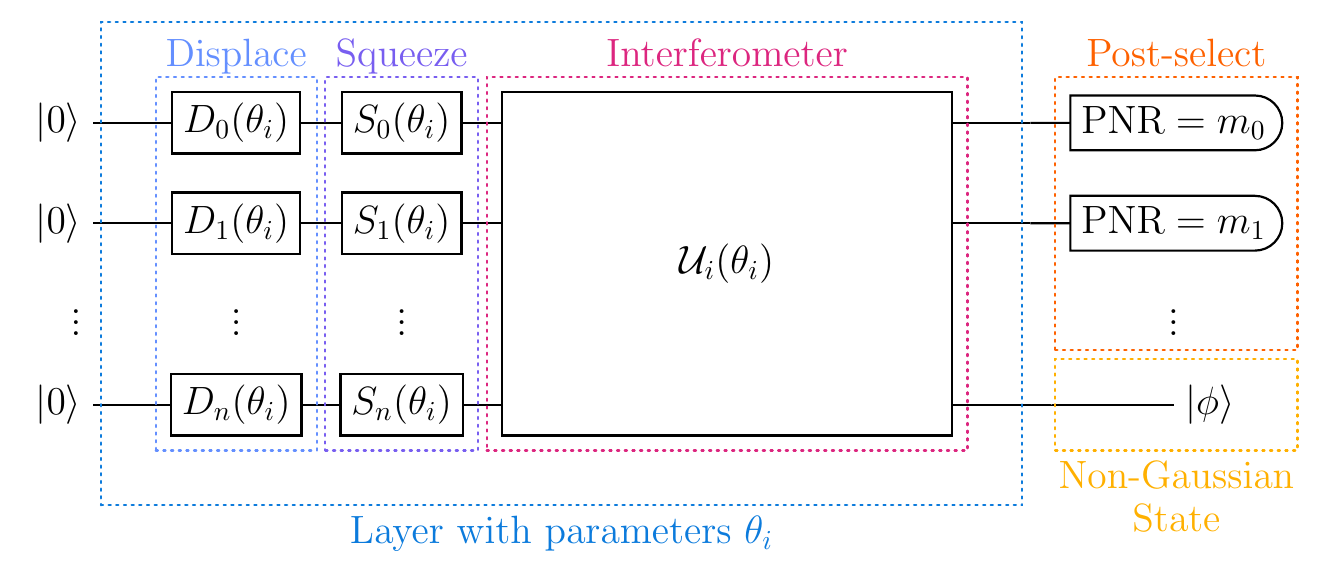}
\caption{}
\label{fig:nngCircuit}
\end{subfigure}
\begin{subfigure}[c]{0.6\textwidth}
\centering
\includegraphics[keepaspectratio, width=\textwidth]{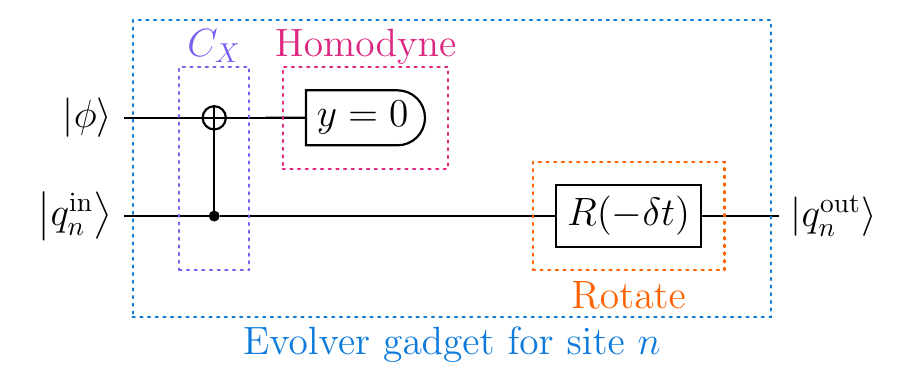}
\caption{}
\label{fig:evolverGadget}
\end{subfigure}
\caption{A measurement-based approach to simulating scalar quantum field theories on a photonic quantum device. In (a), a machine-learning optimised circuit prepares an ancilla state, $\ket{\phi}$,  which encodes the non-Gaussian contributions to the time-evolution operator. This state is used by the evolver-gadget in (b) to deform a target qumode, first entangling it with the ancilla, and then performing a post-selected homodyne measurement. The Gaussian contributions to the time evolution are then implemented through a single rotation gate. The resulting action of the evolver-gadget corresponds to a single Trotterised time step of a quantum state represented on the target qumode. }\label{fig:qumodeEvolutionCircuit}
\end{figure}

Figure~\ref{fig:nngCircuit} shows a schematic of the measurement-based quantum circuit for generating such non-Gaussian states on a single qumode\footnote{By measuring fewer qumodes, this protocol can be extended to prepare multi-qumode non-Gaussian states~\cite{PhysRevA.100.052301}.}. The circuit comprises $N$ qumodes, each of which is initially displaced and squeezed before passing through an interferometer. This routine forms a layer that can be repeated to increase the circuit's ability to generate the desired entangled state on the qumode system. Photon-number-resolving (PNR) measurements are then performed on $N-1$ qumodes, and post-selection on the resulting PNR outcomes induces a non-Gaussian state on the remaining $N$~th qumode. The photonic circuit can be embedded into a classical machine-learning toolchain to train the circuit parameters, $\left\{\theta_i\right\}$ (i.e. the gate parameters of the beamsplitter, squeezing and displacement operations), to generate a non-Gaussian {\it ancilla} state, $\ket{\phi}$, on the $N$~th qumode. 
The desired form of the non-Gaussian ancilla state for the real-time evolution is $\ket{\phi} = {\mathcal U}_V( \hat q')|0\rangle$ where $\hat q'$ represents the quadrature variable of the ancilla qumode, $|\phi\rangle$. That is explicitly we have 
$\langle  q' | \phi\rangle = \langle  q' | \exp (-i \delta t V_I(q') ) |0\rangle $, where $V_I(\hat q')$ is the interacting potential given by Eq.~\eqref{eq:vvv}. It was shown in Ref.~\cite{Abel:2024kuv} that circuits comprising four qumodes were sufficiently expressive to efficiently train the required non-Gaussian states.

Having prepared this non-Gaussian ancilla state, $\ket{\phi}$, we may then employ the \textit{evolver-gadget} from Ref.~\cite{Abel:2024kuv}. The evolver gadget operates in two stages. The first implements the ${\mathcal U}_V$ piece of the evolution by entangling the ancilla qumode with the target qumode and then performing a homodyne measurement on the ancilla, post-selecting on the outcome. The second then applies the ${\mathcal U}_R$ piece of the evolution through a rotation gate applied to the target qumode. This procedure follows the approach outlined in Ref.~\cite{PhysRevA.100.012326}. As Ref.~\cite{Abel:2024kuv} shows, the measurement-induced noise factor can be maximally suppressed by carefully tailoring the non-Gaussian state, $\ket{\phi}$. A schematic of the evolver-gadget is shown in Fig.~\ref{fig:evolverGadget}, where the optional squeezing gate has been removed for clarity. Altogether, the evolver-gadget therefore evolves the system with the evolution operator
\begin{equation}
    {\mathcal U}_{\rm diag} ~=~ e^{-i H_{\rm diag} \delta t} = e^{-iH_0 \delta t} \,{\mathcal U}_V ~+ {\cal O}(\delta t^2),
\end{equation}
where $H_0$ is the SHO Hamiltonian from Eq.~\eqref{eqn:SHO}. 

\begin{figure}[t!]
\centering
\includegraphics[keepaspectratio, width=0.7\textwidth]{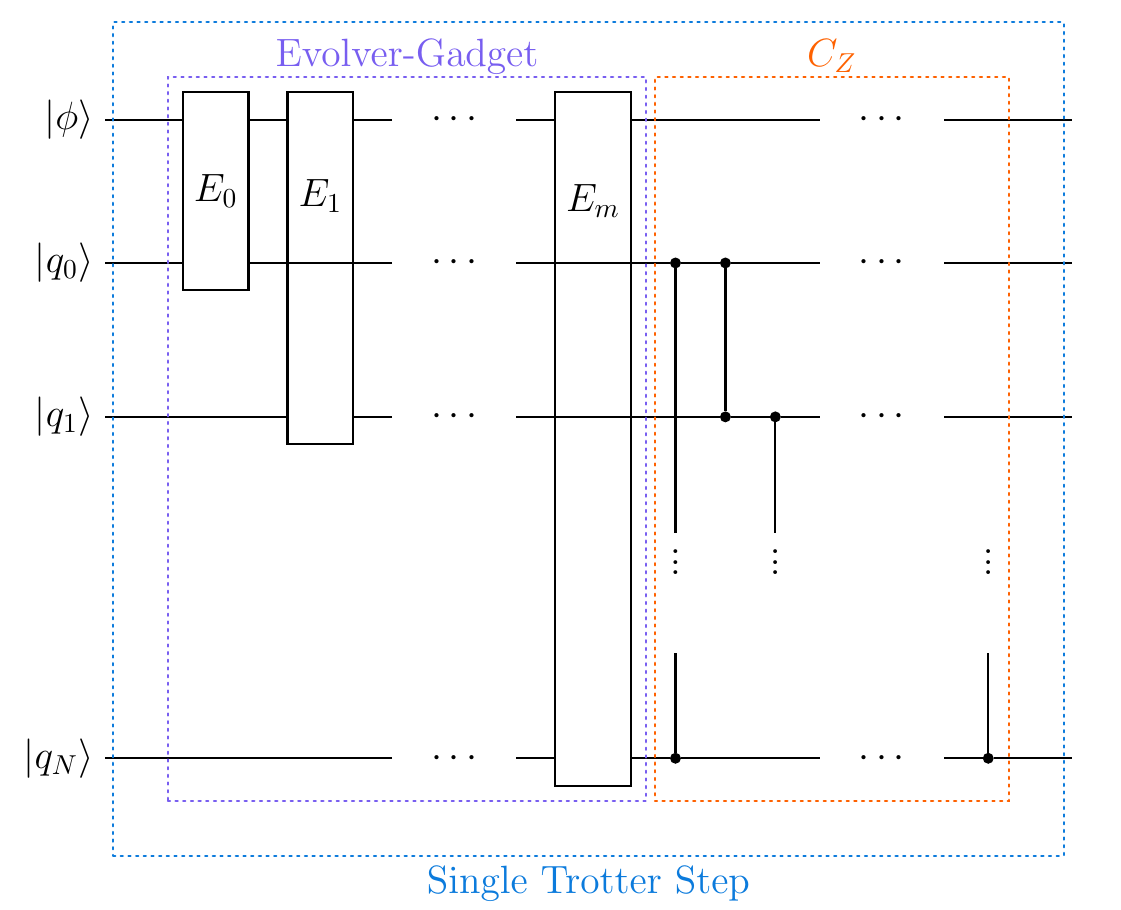}
\caption{A schematic of a quantum circuit that can simulate the real-time evolution of a scalar quantum field theory. The evolver-gadget evolves each qumode under an arbitrary potential prepared using the ancilla state $\ket{\phi}$. Each qumode is then coupled to its nearest neighbour through the series of controlled-Z gates, simulating the hopping-term interaction between qumodes.}
\label{fig:qftCircuit}
\end{figure}

In a single Trotter step, the evolver-gadget must be applied to each qumode in turn to implement the real-time evolution corresponding to the first term, $H_{\rm diag}$, in Eq.~\eqref{eqn:CVQCHam}, and then all the qumodes must be coupled together to reproduce the hopping-term, ${\mathcal U}_{\rm hop}$. This is achieved through a series of controlled-Z gate operations, each of which has exactly the form of a single hopping term in the evolution operator. Indeed, a controlled-Z gate between two qumodes with quadratures $\hat{q}_a$ and $\hat{q}_a$ has the form 
\begin{equation}
    U_\textrm{Z}(a,b,\theta ) ~=~ e^{i \hat{q}_a {\hat q} _b \, \theta }~,
\end{equation}
so that 
\beq \label{eq:hopping}
{\mathcal U}_{\rm hop} ~=~ \prod_{n=0}^{N-1}  U_\textrm{Z}(n,\overline{n+1},\delta t 
)~.
\eeq
Combined with our evolver gadget, the real-time evolution of scalar QFT through a single Trotter step can be implemented directly on a photonic device. A full schematic of the quantum circuit is shown in Fig.~\ref{fig:qftCircuit}. 

\section{Real-time evolution of  (1+1)-dimensional $\lambda\varphi^4$ theory}\label{sec:phi4}

To test the qumode formulation of scalar quantum field theory (QFT) outlined in Sec.~\ref{sec:realTime}, we consider the real-time evolution of field theory in (1+1)-dimensions with interaction potential 
\beq
{\mathscr{V}}_I(\varphi) ~=~ \lambda\varphi^4~.\eeq 
where $\lambda$ is the coupling. 
Following the prescription outlined above, the lattice Hamiltonian has a corresponding effective potential,
\begin{equation}
    H_1 = \frac{1}{a^2} \hat{q}_n^2 + \frac{\lambda}{4!}\hat{q}_n^4~,
\end{equation}
In most of this section we shall be emulating the quantum device using the framework outlined in Sec.~\ref{sec:basics}, in which we work with a lattice of classical modes.  As we shall see this allows for the efficient simulation of hundreds of qumodes. We will then discuss how the same study would be implemented on a full quantum lattice.

\subsection{The two-point function} 

\begin{figure}[!h]
\begin{center}
\begin{subfigure}[c]{0.85\textwidth}
\centering
\includegraphics[keepaspectratio, width=\textwidth]{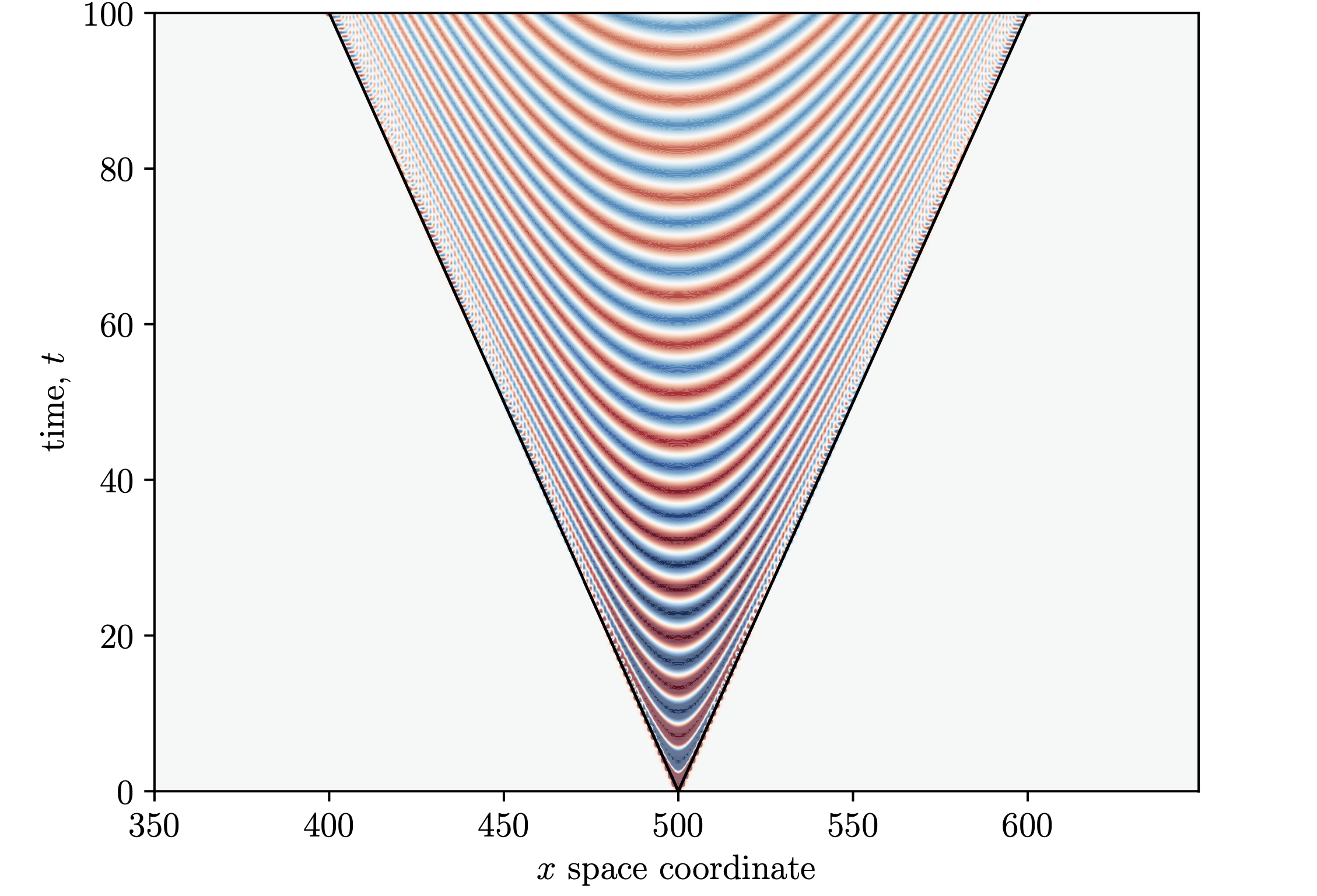}
\caption{}
\label{fig:2ptpert}
\end{subfigure}
\begin{subfigure}[c]{0.85\textwidth}
\centering
\includegraphics[keepaspectratio, width=\textwidth]{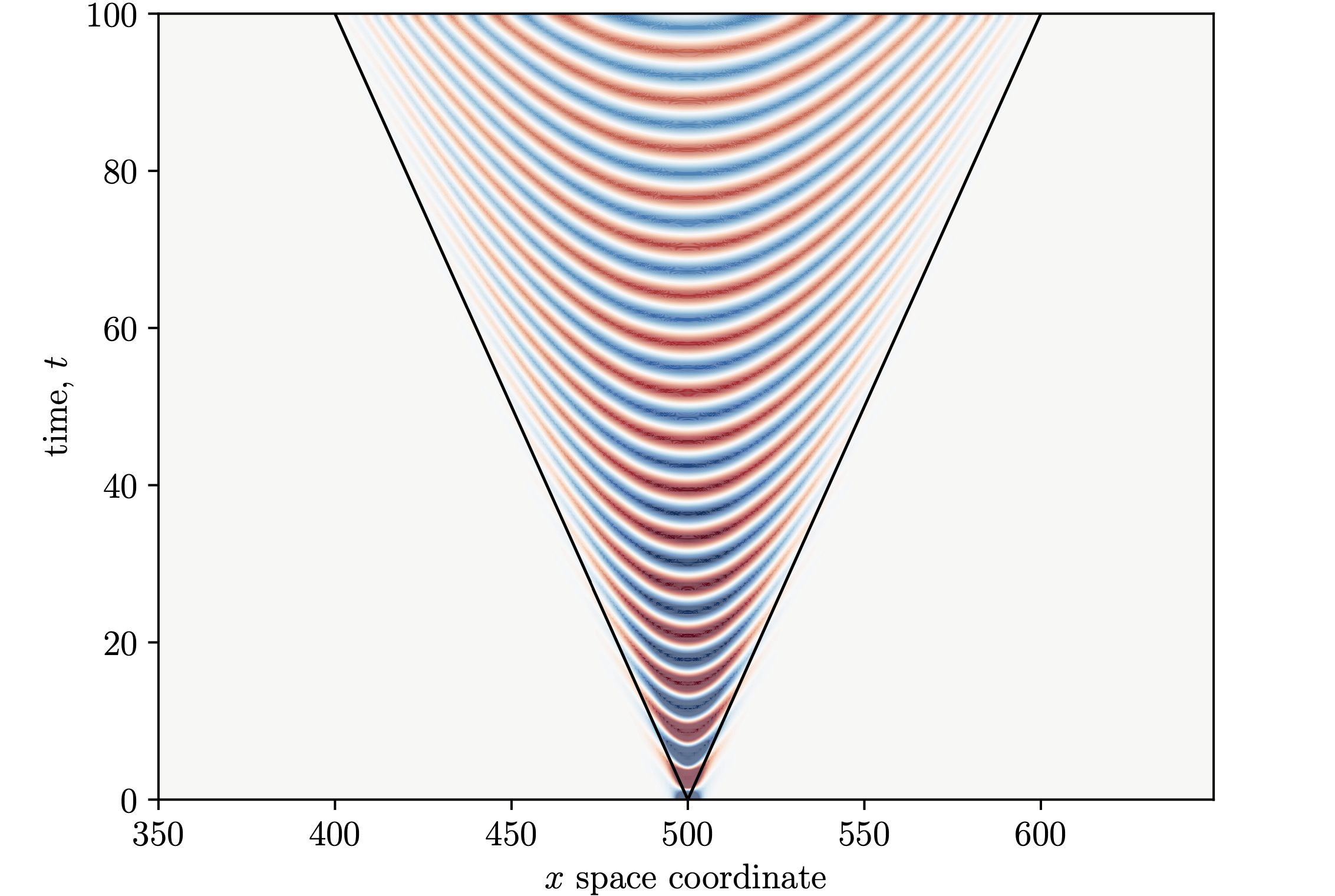}
\caption{}
\label{fig:2ptlat}
\end{subfigure}
\caption{Figure (a) shows the analytical calculation of the two point function $D_R(x-x_0,t)$. Figure (b) shows the field value induced on the Hamiltonian lattice by an approximation to a delta function impulse at $(x,t)=(x_0,0)$ where we take $x_0=Na /2$.}
\label{fig:prop}
\end{center} 
\end{figure}

An important building block for QFTs is the propagator, which encapsulates how particles propagate through spacetime, effectively describing the evolution of quantum states from one point in spacetime to another. Therefore, as a simple initial test of our Hamiltonian qumode lattice formulation, we begin by  simulating the time evolution of the free field in order to reproduce the propagator. 

We set $\lambda =0$, and as we are performing real-time evolution wish to compare our results with the retarded propagator for the scalar theory whose familiar form in momentum space is (in the $i\epsilon $ prescription)  
\begin{equation}
    \widetilde{D}_R = \frac{i}{p^2 - m^2 + i\epsilon p_0 }~,
\end{equation}
where $m^2 = \omega^2$ in the free case. It is a standard but nontrivial exercise to Fourier transform the propagator to position space. In (1+1)-dimensions this yields the closed-form solution in terms of $\tau^2 = (t-t')^2 - (x-x')^2$:
\begin{align}
D_R(t-t',x-x' ) ~=~  \frac{1}{2}
    \Theta ( t-t') \, \Theta (\tau^2)   J_0 (m \tau ) ~,
\end{align}
where $J_0$ is the  Bessel function of order zero, and $\Theta $ is the Heaviside function.  Fig.~\ref{fig:2ptpert}, shows the function $D_R(t-t',x-x' ) $.

In order to probe the two point function we can examine the response of the qumode lattice to a perturbation  of the field \(\phi\) 
  at \((x=0, t=0)\) with a delta function. That is we impose an initial condition on the field and its time derivative at \(t=0\):
\begin{equation}
    \phi(x, 0) = \delta(x), \quad \partial_t \phi(x, 0) = 0. \label{eq:2}
\end{equation}
The solution to the initial value problem \eqref{eq:2} at later times \(t > 0\) is given by
\begin{align}
    \phi(x, t) &~=~ \int dx' \, D_R(x - x', t) \, \phi(x', 0)\notag \\
   &~=~ D_R(x, t)~, 
   \label{eq:4}
\end{align}
and thus we expect the field value to reproduce the propagator directly. 

\begin{figure}[t]
    \centering
    \includegraphics[width=0.8\linewidth]{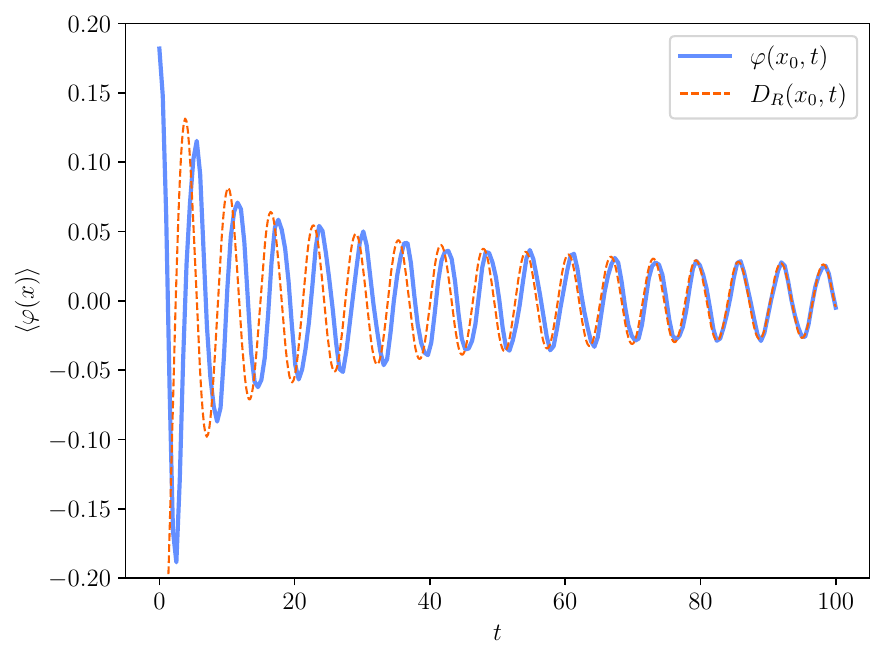}
    \caption{Comparison of the field $\varphi(x_0,t)$ (solid line) generated by a $\delta $-function impulse at $t=0$, with the propagator $D_R(x_0,t)$ (dashed line), where we take $x_0 = Na/2$. Here we take a Trotter step of $\delta t = 0.01$. }
    \label{fig:slice}
\end{figure}

To test this, we begin with an initial state in which the field expectation value approximates a delta function situated at $x=aN/2$. The simplest way to achieve this is to displace the Gaussian ground state of the single qumode at $x=aN/2$. (On the actual device this could be enacted with a displacement gate.) The size of the displacement is of course limited, so this can be thought of as a `top-hat' approximation to the delta function.

Figure~\ref{fig:2ptlat} shows the resulting time-evolution of the field  calculated using the classical emulation of the CVQC approach from Sec.~\ref{sec:basics}. The quantum simulation shows remarkable agreement with the exact calculation, exhibiting the required light-cone structure represented by the black lines, and the oscillatory nature of the field within the light-cone, represented by the alternating colours of the density plot. 
We also show in Figure~\ref{fig:slice} the field value from an initial impulse at $(x,t)=(x_0,0)$ with the function $D_R(x_0,t)$, at the point  $x_0 = Na/2$.

In both cases we see a slight discrepancy between the classical and quantum simulations for field values close to the light cone. This is due to the inability to  construct an exact delta function for the initial state, because the delta function cannot be narrower than the lattice spacing $a$. However, away from the light-cone in the deep time-like region the evolution on the lattice shows good agreement with the analytic result. Overall, Figs.~\ref{fig:prop} and \ref{fig:slice} demonstrate the CVQC framework's ability to efficiently capture physical effects with a very simple implementation. Having calculated the propagator, let us now turn to scattering wave packets on the lattice. 

\subsection{Initial state preparation on the classical mode lattice}\label{sec:initialState}

\begin{figure}[t]
    \centering
    \includegraphics[width=\textwidth]{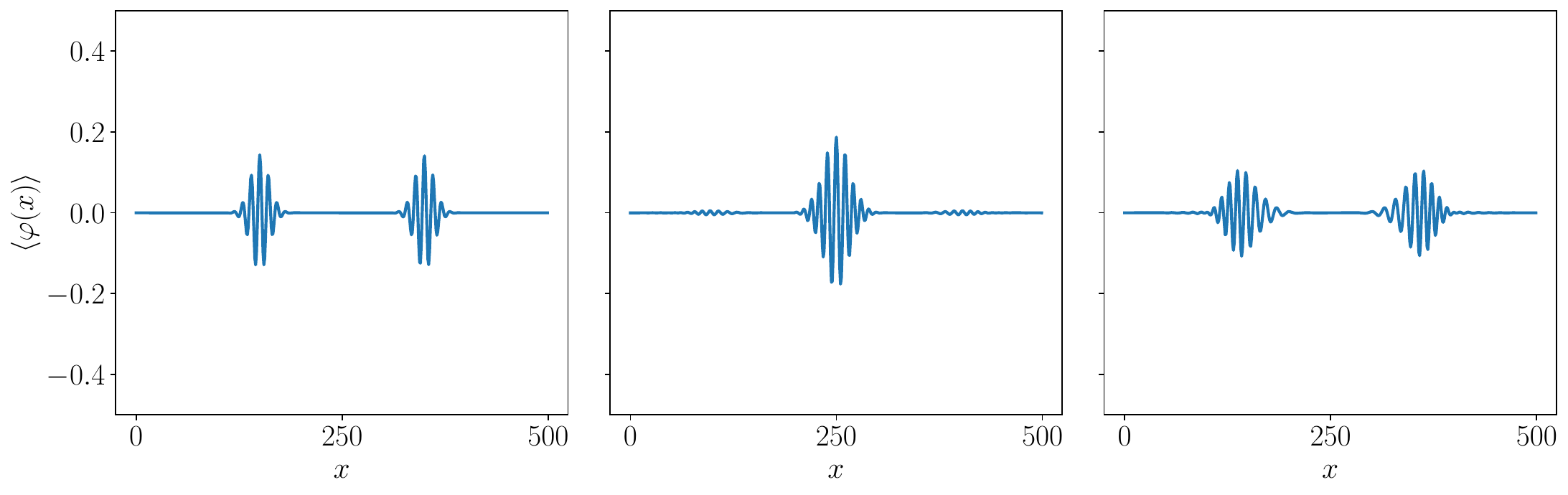}
    \caption{Scattering wavepackets simulated on the lattice. The left panel shows the initial state Gaussian wavepackets, the middle shows the wavepackets as they collide and the right panel shows the post-collision wave packets.}
    \label{fig:wavepackets}
\end{figure}

To perform scattering experiments, it is essential to prepare well-defined initial states. As discussed in Sec.~\ref{sec:basics}, preparing specific field configurations of the QFT is non-trivial, and has been shown to be a QMA-complete problem for local Hamiltonians~\cite{doi:10.1137/S0097539704445226}. 
To validate our framework, we wish to simulate a scattering process classically using the method outlined in Sec.~\ref{sec:basics}, constructing two wavepackets and simulating the real-time evolution under the scalar QFT Hamiltonian from Eq.~\eqref{eqn:CVQCHam}. As per Eq.~\eqref{eq:initPsi} and the discussion in Sec.~\ref{sec:basics}, these will be constructed from momentum excitations of the vacuum by modes that are non-relativistic in order to obtain a good approximation of the QFT whilst remaining in the qumode basis. Thus, we may  build a Gaussian wavepacket centred at position $\bar{x}$ with momentum $\bar{k}$ and with a spread in the momentum of $\sigma$~\cite{kitaev2008wavefunction}. We will maintain a non-relativistic approximation by imposing both $\bar{k}\ll \omega$ and $\sigma \ll \omega$. Therefore, utilising the  direct identification of field operator with quadrature variable given in Eq.~\eqref{eq:phi_to_q}, we may take 
\begin{equation}\label{eqn:gaussianWavepacket}
   \ket{\psi_n} ~=~\left( A^0 +  \sum^{N-1}_{\alpha=0} \frac{1}{\sqrt{\mathcal{N}_\alpha^{\phantom{2}}}} \exp\left(- \frac{(k_\alpha - \bar{k})^2}{2\sigma^2} \right) \exp \left( i k_\alpha (x_n - \bar{x}) \right) \hat a ^\dagger_n  \right) \ket{0}~,
\end{equation}
where $k_\alpha$ is defined in Eq.~\eqref{eq:NinftyK}, and $x_n=a n$, and where  $\mathcal{N}_\alpha$ is the normalisation factor for the field components, while the amplitude of the zeroth Fock mode of the qumodes, $A^0$, is adjusted to normalise them:
\begin{align}
    \mathcal{N}_\alpha ~&=~ { 2\omega_\alpha N }\nonumber \\
    |A^0|^2 ~&=~ 1 - 
    \sum_{\alpha=0}^{N-1}
    \frac{1}{\sqrt{\mathcal{N}_\alpha }}\, 
     e^{-\frac{4\pi^2 \alpha^2 }{L^2\sigma^2}} ~.
\end{align}
In the classical simulation we can of course simply initiate the lattice with the Fock amplitudes of each qumode set to have the requisite amplitudes. This is also possible by using the `Fock backend' within available simulators such as {\tt Strawberryfields} \cite{Killoran2019strawberryfields,Bromley_2020}.  Figure~\ref{fig:wavepackets} shows an illustrative example of such wavepackets prepared in the qumode basis, and shows how they evolve over time, where here and henceforth we set $a=1$

\subsection{Scattering in (1+1)-dimensional $\varphi^4$ theory}

We present the results from $\varphi^4$ theory scattering experiments in (1+1)-dimensions, focusing on the interaction dynamics of field configurations and energy densities. Using the classical emulation of a CVQC device outlined in Sec.~\ref{sec:basics}, we simulate the real-time evolution of wavepackets under various parameters, including different mass and coupling strengths. The system we will consider comprises 500 simulated qumodes, corresponding to a lattice of 500 sites. The quadrature variable for each qumode has been discretised as in Eq.~\eqref{eq:discretise}, where a discretisation of $M=200$ has been chosen. 

\begin{figure}[t]
\centering
\includegraphics[keepaspectratio, width=\textwidth]{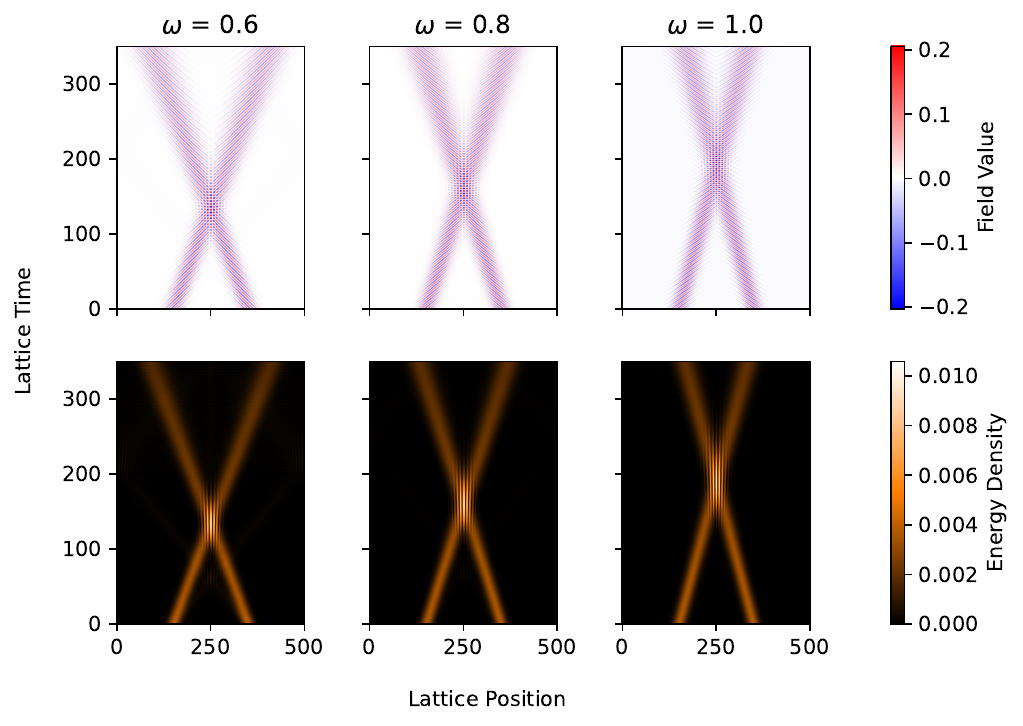}
\caption{Scattering in $\varphi^4$ theory with $\lambda=0.2$ and varying masses.}
\label{fig:mass}
\end{figure}

To carry out the scattering simulations, we initialise the system with two Gaussian wavepackets of the form shown in Eq.~\eqref{eqn:gaussianWavepacket} centred around positions $x=150$ and $x=350$, with a spread of $\sigma=0.09$. Each wavepacket is initialised with an initial momentum of $k_\alpha = 0.3$ to respect the non-relativistic requirement of the classical emulation, thus retaining a good approximation of the QFT. The Trotterised real-time evolution has been simulated for a total simulation time of $T=350$ in lattice units with a time step of $\delta t= 0.01$. The method is shown in Fig.~\ref{fig:testplot} to be resilient to Trotter error up to equivalent times for the evolution of a single qumode. Therefore, we expect the full QFT simulation to be similarly resilient, with Trotter errors remaining small up to time $T$. 

We performed two sets of simulations to test the framework. The first reveals the effects of changing the mass of the particles in the scattering process. Figure~\ref{fig:mass} shows how the dynamics of the process change for different masses, $\omega=0.6$, $0.8$~and~$1.0$, with a fixed coupling of $\lambda=0.2$, close to the free theory. The top row in the figure shows the field value and the bottom row shows the energy density of the system. One can observe that increasing the mass slows the propagation of the wavepackets and shifts the interaction vertex in time, delaying the interaction. Furthermore, we see that the wavepackets become more localised as the mass increases, and the oscillations reduced in the field value. 

\begin{figure}[t]
\centering
\includegraphics[keepaspectratio, width=\textwidth]{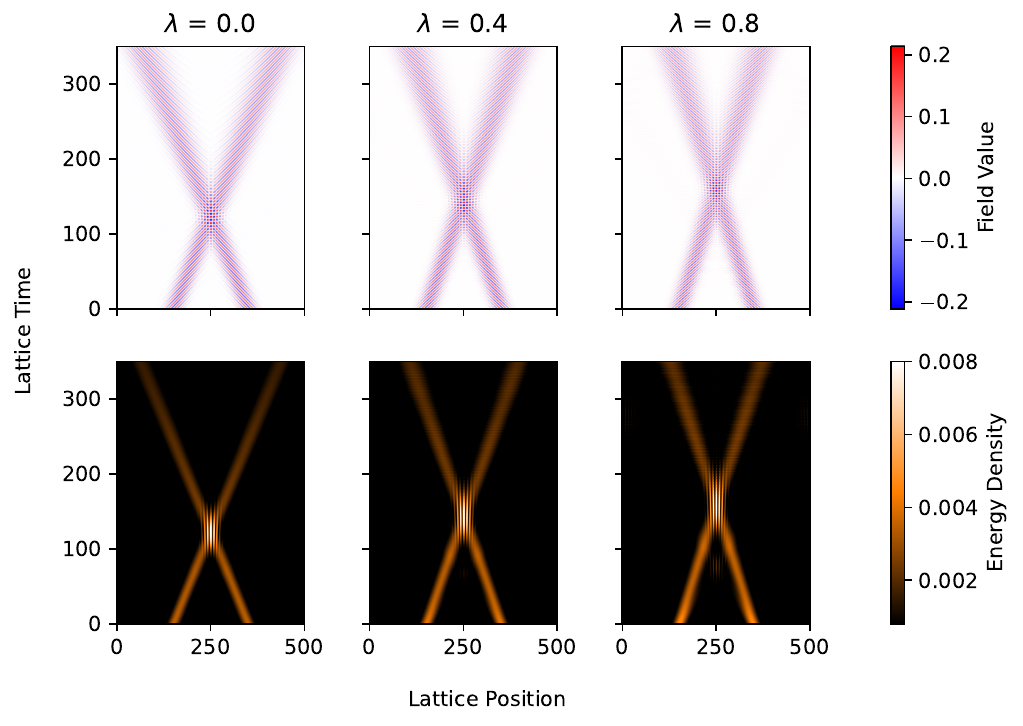}
\caption{Scattering in $\varphi^4$ theory with $\omega=0.6$ and varying coupling.}
\label{fig:coupling}
\end{figure}

The second scattering experiment examines the effect of increasing the coupling. For this scenario, we fix mass to $\omega=0.6$ and increase the coupling for values of $\lambda=0.0$, $0.4$ and $0.8$. From the results shown in Fig.~\ref{fig:coupling}, we see that the interaction vertex is once again shifted in time, with larger couplings slowing the interaction. This is consistent with what one expects from a repulsive theory such as $\lambda \varphi^4$ with $\lambda > 0$. Furthermore, the simulations indicate that larger $\lambda$ values lead to significant non-linearities in the field dynamics. These non-linear effects manifest as pronounced deformations in the field evolution and enhanced energy dispersion across the lattice. 

These simulations show the critical role of both the mass and the coupling strength in determining the scattering dynamics within the $\lambda \varphi^4$ theory. The mass governs the speed of wavepacket propagation, while the coupling strength introduces non-linear effects that significantly alter the field’s evolution and energy distribution. 

Both experiments were simulated on an Apple M2 Pro chip with 16GB of RAM. Simulating 500 qumodes, each with a field-discretisation of $M=200$, for $T=350$ lattice time required approximately one hour per run, illustrating the efficiency of the classical emulator described in Sec.~\ref{sec:basics}. 

\section{Aspects of the quantum qumode lattice}\label{sec:aspects}

Having validated the qumode lattice's ability to simulate the real-time evolution of quantum field theories (QFTs) in Sec.~\ref{sec:phi4} using the classical simulation of the system, we now turn our attention to those aspects that are unique to implementation on a genuine quantum qumode lattice. Specifically, in this section we will outline and identify the advantages to be gained by simulating QFTs on continuous-variable quantum devices. 

\subsection{Adiabatic state preparation}

We first consider aspects of state preparation. 
To perform scattering calculations in real-time scalar field theories, one can be guided by the prescription of Ref.~\cite{10.5555/2685155.2685163}. First, one prepares the lattice of uncoupled qumodes in the ground state. This ground state is trivially the state in which each individual qumode is in the SHO ground state, and is the same vacuum that we used for our non-relativistic approximation above. The lattice of uncoupled qumodes can then be excited to construct Gaussian wavepackets~\cite{kitaev2008wavefunction}. A Gaussian wavepacket on the uncoupled lattice we refer to as a 
``proto-wavepacket''. 

Implementing the proto-wavepackets in the CVQC framework can be done with a straightforward series of displacement gates. As shown in Eq.~\eqref{eq:phi_to_q}, the field value at lattice site $n$ directly corresponds to the quadrature value of qumode $n$. The expectation value of the quadrature variable, and by extension, the field value, can be changed via a displacement operation on the qumode. Therefore, to initialise the proto-wavepacket on the quantum device, each qumode is simply displaced accordingly to reflect the amplitudes of the wavepacket. If we follow the approach of Sec.~\ref{sec:phi4}, we note that the completely uncoupled qumode lattice is equivalent to sending $a\to \infty$. It is a simple exercise to verify that the proto-wavepackets that we put on the lattice initially should then have the following field VEVs:
\begin{equation}\label{eqn:gaussianWavepacket2}
    \langle{\varphi(x_n, t=0)}\rangle  ~=~ \frac{A_0}{\omega N} 
    \sum_{\alpha = 0}^{N-1} 
\exp\left(- \frac{(k_\alpha - \bar{k})^2}{2\sigma^2} \right) \cos \left( k_\alpha (x_n - \bar{x}) \right)~.
\end{equation}
Note that only the prefactor ${\cal N}_\alpha$ scales with $a$, becoming universal in the large $a$ limit according to Eq.~\eqref{eqn:omegas}, and sending ${\cal N}_\alpha\to \sqrt {\omega N}$, while the other terms in the wavepacket are unaffected. In particular our earlier rotations between the qumode lattice basis and the QFT momentum basis are not dependent on $a$. Thus, to create the proto-wavepacket we can simply displace the gates according the VEV in Eq.~\eqref{eqn:gaussianWavepacket2}. Note that the resulting field would be entirely real, implying that what is being created is the sum of a left-moving and right-moving component. To initialise proto-wavepackets in such configurations would effectively be to start them in the middle panel of Figure~\ref{fig:wavepackets}. Half of the wavepackets would then fly off away from the collision point, but they could be safely ignored, as long as they do not interfere with the collision. (In order to instead produce just isolated left- or right-moving modes, phases could be introduced by using a set of carefully tuned gates for each qumode, at the price of significant complication.)

The next step after preparing the proto-wavepackets is to move to the interacting QFT. We have already discussed the difference between the qumode lattice vacuum and the QFT vacuum, and seen that the addition of hopping terms in principle completely changes the nature of the vacuum to that given in Eq.~\eqref{eq:trueVac}, which has important implications for ultra-relativistic modes. In addition, the interactions in the theory will also change the nature of the vacuum from that of the free theory. Thus we need to effectively dial up the hopping terms that couple the qumodes to each other, as well as the non-linear interaction terms in the potential, $V_I$. 
A key advantage of the quantum qumode lattice is that this transition to the interacting QFT can be achieved via an adaptation of the methods of adiabatic state preparation~\cite{10.5555/2685155.2685163, farhi2000quantumcomputationadiabaticevolution, doi:10.1126/science.1217069, RevModPhys.90.015002, 10.1063/1.2798382, 10.1063/1.4748968}, which is a well-established method for approximating eigenstates in complex quantum systems. This approach employs the adiabatic theorem, which states that eigenstates in a particular energy level of an initial Hamiltonian, $H_i$, will remain as instantaneous energy eigenstates under a deformation if the evolution is applied sufficiently slowly. For the qumode lattice, this means that we can start in the ground state of the qumode system and adiabatically evolve towards the ground state of the full QFT, by slowly turning on both the hopping terms and the interaction terms in the Hamiltonian at the same time. Moreover Gaussian proto-wavepackets initialised on the decoupled qumode system would yield corresponding Gaussian wavepackets in the QFT. As in Ref.~\cite{10.5555/2685155.2685163}, if performed for a time $\tau$ which is sufficiently long to be adiabatic, but sufficiently short to be completed before the scattering occurs, this process will yield Gaussian wavepackets scattering in the true QFT vacuum on the device. In other words, the proto-wavepackets that were introduced initially on the qumode lattice, being composed from a superposition of energy eigenstates, will simply remain as wavepackets in the system as it adiabatically transitions to the full theory.

\subsection{Quantum resource scaling}

Perhaps the most significant advantage of continuous-variable quantum computing (CVQC) over discrete-variable quantum computing (DVQC) for simulating quantum field theories (QFTs) via lattice models is the ability to maintain the continuous nature of the field. In DVQC approaches, the field value at each lattice site must be digitised, i.e. the field value is discretised and encoded on a set of qubits, $n_\phi$. Consequently, each lattice site in the model carries an overhead of $n_\phi$-qubits per site. Reference~\cite{10.5555/2685155.2685163} estimates that achieving a tolerable error requires $n_\phi= \mathcal{O} (10)$, thus  an order of ten-thousand qubits are needed to simulate scattering events in (1+1)-dimensional scalar field theories with percent-level accuracy. Reducing the number of qubits increases the discretisation error, thus marking the trade-off between accuracy and quantum resource cost as a fundamental limitation of DVQC lattice approaches\footnote{It should be noted that non-lattice approaches have been shown simulate non-equilibrium physics using small numbers of qubits, such as Hamiltonian Truncation approaches to quantum quenches~\cite{Ingoldby:2024fcy}.}.

In stark contrast, the quantum qumode lattice removes the need to digitise the field value, instead encoding the field value as a continuous variable on the qumode expectation value. This dramatically reduces the quantum resource requirement by a factor of $n_\phi$, which scale linearly with the number of lattice sites. As demonstrated in Sec.~\ref{sec:phi4}, scattering events can be simulated efficiently using just 500 qumodes, two orders of magnitude fewer qumodes than the number of qubits required for the DVQC approach. 

Furthermore, the quantum qumode lattice method outlined in Sec.~\ref{sec:realTime} has an efficient gate depth for a single Trotter step, growing linearly with the number of lattice sites as it requires only nearest neighbour multi-qumode operations. Consequently, CVQC provides a quantum resource-efficient alternative with experimentally realisable gate operations on a feasible number of qumodes, making it a promising framework for near-term quantum simulations of QFTs.

\section{Summary and Conclusions}\label{sec:conclusions}

In this study, we have presented a framework for the real-time simulation of quantum field theories (QFTs) using continuous-variable quantum computing (CVQC). Focusing on $\varphi^4$ scalar field theory in $(1+1)$-dimensions, we have proposed a Hamiltonian formalism, representing the field on a spatial lattice with field values encoded as quantum harmonic oscillators or qumodes. This mapping takes full advantage of the CVQC's inherent ability to handle continuous-variable systems, eliminating the need for field digitisation and enabling efficient navigation of large Hilbert spaces. 

The quantum algorithm simulates the Trotterised time-evolution of a scalar field theory using experimentally realisable quantum gate operations, enhanced by a classical machine learning tool-chain. Developing methods for preparing initial quantum states corresponding to specific field configurations is critical to successfully simulating the real-time evolution of QFTs. We outline how this can be achieved efficiently using simple displacement gate operations and adiabatically evolving the system by slowly ramping-up the interaction terms between qumodes. 

To validate the CVQC method, we utilised a classical quantum emulation technique to simulate the real-time evolution. In the classical case, the initial state is prepared in as a non-relativistic perturbation of the qumode device vacuum, and has been shown to be a good approximation of a scattering state in the QFT basis. These prepared states were evolved under the $\varphi^4$ Hamiltonian, allowing one to capture the dynamics of field interactions and compute key observables such as two-point correlation functions and transition matrix elements. The results of these simulations validate the CVQC framework, demonstrating its capacity to accurately replicate known analytical solutions in the free-field regime and to extend beyond them to interacting theories.

The free-field simulations provided an important benchmark for the CVQC methodology. The emerging two-point correlation functions compare exceedingly well with analytical solutions, affirming the reliability of our approach. 

Our findings for the interacting theory were in agreement with expectations for how the mass and coupling strength of the field influence scattering dynamics. As the mass parameter increases, the mobility of the field excitations decreases, dispersion increases, and the energy density becomes more localised and oscillation amplitudes reduced. In contrast, increasing the coupling strength introduces non-linearities in the field evolution, leading to significant deformations in the dynamics and broader redistribution of energy across the lattice. These results show the sensitivity of scalar field behaviour to its fundamental parameters and display the correct interplay between mass, coupling, and field evolution.

An important aspect of this approach is that the continuous nature of CVQC leads to a linear scaling in the required quantum resources as the number of lattice sites grows. This is because the field remains continuous and does not require digitisation. This scalability, coupled with the low-decoherence properties of photonic platforms, positions CVQC as an ideal tool for studying quantum fields and other continuous-variable systems.

Our approach to integrating experimentally realisable non-Gaussian operations into the simulation of QFTs on photonic devices lays the groundwork for exploring more complex QFTs using continuous variables, establishing CVQC as a potentially powerful and versatile framework for investigating non-equilibrium quantum phenomena, dynamic phase transitions, and the intricate behaviour of quantum fields in the near term. Advancements in photonic quantum hardware will be pivotal in unlocking this potential, enabling the study of increasingly complex and computationally demanding quantum systems.

\vspace{0.5cm}
\noindent{\textit{\textbf{Acknowledgments} We would like to thank James Ingoldby and Thomas Stone for valuable discussions.}

\bibliographystyle{inspire}
\bibliography{refs}{}

\end{document}